# Novel solid-phase epitaxy for multi-component materials with extremely high vapor pressure elements: An application to $KFe_2As_2$

Authors:

Taisuke Hatakeyama,[1] Hikaru Sato,[1] Hidenori Hiramatsu,[1,2*] Toshio Kamiya,[1,2] and Hideo Hosono[1,2]

Affiliations:

[1] Laboratory for Materials and Structures, Institute of Innovative Research, Tokyo Institute of Technology, Mailbox R3-1, 4259 Nagatsuta-cho, Midori-ku, Yokohama 226-8503, Japan

[2] Materials Research Center for Element Strategy, Tokyo Institute of Technology, Mailbox SE-6, 4259 Nagatsuta-cho, Midori-ku, Yokohama 226-8503, Japan

[*] E-mail: h-hirama@lucid.msl.titech.ac.jp

Abstract

We propose a novel solid-phase epitaxy technique applicable to high annealing temperatures up to 1000 °C without re-vaporization of alkali metal elements with high vapor pressures. This technique is demonstrated through the successful growth of high-quality $KFe_2As_2$ epitaxial films. The key factors are employing a custom-designed alumina vessel/cover and sealing it in a stainless tube with a large amount of atmospheric $KFe_2As_2$ powder in tightly closed sample spaces. This technique can also be effective for other materials composed of elements with very high vapor pressures, such as alkali metals, and can lead to the realization of spintronics devices in the future using $KFe_2As_2$.

In 2012, Pandey et al. [1] theoretically predicted that $KFe_2As_2$, a superconductor with a critical temperature of 3 – 4 K, [2,3] might exhibit a large spin Hall conductivity (SHC) comparable to that of Pt, which has an SHC $10^4$ times larger than those of conventional semiconductors. [4] The detection of the SHC in $KFe_2As_2$ is limited by its small spin diffusion length (roughly a few tens of nanometers) and requires a sample size (including thickness) much smaller than this value. Therefore, bulk single crystals cannot be used for this purpose, and high-quality nanometer-scale single-domain thin films are desired. For this reason, we attempted to fabricate high-quality $KFe_2As_2$ epitaxial films.

However, the growth of $KFe_2As_2$ films is challenging mainly because of their major constituent, K. It is well known that alkali metals, such as Li, Na, and K, are chemically active and can easily react with water/oxygen in air. This can be overcome if the sample is handled in a dry and inert glove box. Another issue is the extremely high vapor pressures of these elements. For example, the sublimation / melting temperatures at 1 Pa ($T_{1Pa}$) for K and Na are 200.2 and 280.6 °C, respectively. [5] Although Zn and Ga are known to evaporate easily during vacuum deposition at somewhat high temperatures, their $T_{1Pa}$ are 337 °C and 1037 °C, [5] respectively, which are much higher than the $T_{1Pa}$ of the alkali metals. The growth of high-quality epitaxial films requires high temperatures, which, however, will consequently evaporate these high vapor pressure elements. Therefore, in spite of the rather high $T_{1Pa}$ of Zn and Ga, we cannot grow high-quality epitaxial films composed of these elements [e.g., $InGaO_3(ZnO)_m$, $m$ is integer) [6]] through direct deposition and need to employ special techniques, such as conventional or reactive solid-phase epitaxy.

This issue is much more serious for $KFe_2As_2$, because even low-temperature deposition including room temperature causes serious chemical composition deviation, producing an alkali metal-poor film. In 2014, we solved this issue by post-deposition thermal annealing in an evacuated silica-glass tube combined with a pulsed laser deposition (PLD),[7] in which the separation of the film deposition and thermal crystallization processes was the key to growing the target films, as also reported in Refs. 6, 8, and 9. The obtained $KFe_2As_2$ films, however, were not epitaxial but *c*-axis orientated films without in-plane orientation, due to the limitation of a maximum annealing temperature of 700 °C. The same temperature was used for growing K-doped $BaFe_2As_2$ films[10] using the conventional annealing method sealed in an evacuated silica-glass tube. When we raised the annealing temperature to above 700 °C, the films decomposed into $Fe_2As$, $FeAs$, and $Fe$. This indicated that the gas-tightness of this synthesis condition was poor at above 700 °C for K and the alkali metal component K did not remain in the films.

In this study, we developed a novel solid-phase epitaxy technique using a custom-made alumina vessel, which realized a high annealing temperature of 1000 °C without vaporization of K and succeeded in obtaining high-quality heteroepitaxial $KFe_2As_2$ thin films on MgO single crystals.

Thin films with 100–200 nm thicknesses were grown on (001)-oriented MgO single-crystal substrates (area: 10 × 10 $mm^2$, thickness: 0.5 mm). We employed $(La,Sr)(Al,Ta)O_3$ (LSAT) single-crystal substrates in a previous study, where the $KFe_2As_2$ films reacted with the LSAT substrates even at 700 °C,[7] similar to an isostructural arsenide $BaFe_2As_2$,[11,12] leading to low-quality films. Therefore, we employed MgO substrates in this study as we had confirmed that other isostructural arsenide films, such as BaFe$_2$As$_2$, did not react with MgO at even higher growth temperatures (800 – 1050 °C). [13,14] The MgO substrates were annealed at 1000 °C for 30 min in ambient atmosphere to remove surface contaminants and improve the surface morphology prior to film deposition. The in-plane lattice mismatch between KFe$_2$As$_2$ (*a*-axis lattice parameter $a$ = 0.3841 nm, Ref. 2) and MgO ($a$ = 0.421 nm, +9%) is larger than that between KFe$_2$As$_2$ and LSAT ($a/2$ = 0.387 nm, +0.8%) used in the previous study. [7] PLD in vacuum at room temperature using a KrF excimer laser (wavelength: 248 nm, repetition rate: 10 Hz) was employed as a chemical transfer technique. We used a K-rich KFe$_2$As$_2$ bulk polycrystal as the PLD target, which was synthesized from a K-rich starting mixture at a molar ratio of K$_3$As:Fe$_2$As = 1 : 1. The fabrication conditions of the KFe$_2$As$_2$ powder / PLD target and the process parameters of PLD are reported in Ref. 7. These samples were subject to a newly developed solid-phase epitaxy technique, which is illustrated in Fig. 1 and will be described later.

The crystalline phase and quality of the films were determined by X-ray diffraction [XRD, source: Cu K$\alpha_1$ monochromated by Ge(220)]. The fluctuations of the crystallite orientations were characterized using the rocking curves of out-of-plane diffractions ($2\theta$-fixed $\omega$ scans). The in-plane orientation of the films was investigated using a pole figure geometry. All of the XRD measurements were performed using an O-ring sealed sample carrier filled with dry Ar gas to prevent degradation of the samples during measurements.

Figure 1 summarizes the newly developed solid-phase epitaxy process for materials composed of alkali metals with high vapor pressures. At first, an amorphous KFe$_2$As$_2$ layer was deposited on an MgO (100) substrate by PLD using the K-rich KFe$_2$As$_2$ target, as reported in Ref. 7 [Fig. 1(a)]. The resulting film was then transferred directly from the PLD preparation chamber to a dry and inert glove box, and covered with a fresh MgO plate [Fig. 1(b)]. The film was tightly covered with a large amount of stoichiometric $KFe_2As_2$ powder in a custom-designed alumina vessel, in order to compensate the high vapor pressure of K during the high temperature annealing process at 1000 °C. The capping MgO plate is effective in preventing evaporation of the film constituents during thermal annealing as well as chemical reaction between the film surface and covering $KFe_2As_2$ powder. The alumina vessel has two deep spaces with a lateral size of 10.2 × 10.2 $mm^2$ (i.e., almost the same lateral size as that of the MgO substrate) and a depth of 5 mm to fill sufficient $KFe_2As_2$ powder. Then the alumina vessel was covered with a custom-made alumina cover. Both sides of the vessel and the cover were tightly clenched by commercially available bolts and nuts made of alumina. The drawings of the custom-designed alumina vessel and cover are shown in Fig. S1 in the online supplementary data at http://stacks.iop.org/APEX/9/055505/mmedia. This tightly sealed alumina vessel was set in a stainless tube filled with dry Ar. Both ends of the stainless tube were tightly clenched by stainless nuts and gaskets to keep it gas-tight during high temperature annealing. In the previous study,[7] we employed a silica-glass tube to keep the sample and vessel in an evacuated atmosphere. This method required a high temperature heating process to seal the glass tube while keeping the sample at lower temperatures (close to room temperature), which was technically difficult. In the present process, however, we can easily perform the sealing in a glove box without heating. The present technique does not require any special experimental skills and can be performed by anyone and applicable for any materials that will easily decompose or react when heated. All the processes in Fig. 1(b) were performed in a dry and inert glove box. This tight sealing method prevents vaporization of K during high temperature annealing even at 1000 °C for 30 min [Fig. 1(c) left]. The maximum available temperature by this technique is 1000 °C, because annealing at above 1000 °C causes a re-vaporization of K, similar to the previously reported technique at 800 °C in Ref. 7. After annealing using the present technique, the stainless nuts were sintered to the stainless tube and could not be loosened. Thus, we cut open the tube by a tube cutter and then picked out the sample in the glove box [see Fig. 1(c), right photograph].

Figure 2 shows the XRD patterns of the $KFe_2As_2$ films grown at 1000 °C through the new solid-phase epitaxy technique. The out-of-plane $\omega$-coupled $2\theta$ synchronous scan pattern [Fig. 2(a)] indicates that the obtained films are strongly $c$-axis oriented. The full width at half maximum (FWHM) of the 002 diffraction rocking curve [Fig. 2(b)] is 0.06°, which is nearly one order smaller than that obtained using the previous technique (0.4°).[7] Figure 2(c) shows the in-plane $\phi$-coupled $2\theta\chi$ synchronous scan pattern, indicating that the films are oriented in-plane as well as out-of-plane. Because the $a$-axis lattice parameter of the film is close to that of MgO, the peaks from the $KFe_2As_2$ film and MgO substrate could not be resolved. Therefore, we performed pole figure measurements of the asymmetric 103 diffraction [see Fig. 2(d)] to confirm the in-plane symmetry of the film. We observed a clear four-fold symmetric diffraction due to the tetragonal symmetry of $KFe_2As_2$ at $\Psi = 50°$ and every $\phi = 90°$. These results indicate that the $KFe_2As_2$ films grew heteroepitaxially on MgO (001) single-crystal substrates with an epitaxial relationship: [001] $KFe_2As_2$ ∥ [001] MgO for out-of-plane and [100] $KFe_2As_2$ ∥ [100] MgO for in-plane. This result is significantly different from that of the previous study,[7] in which $c$-axis oriented films were obtained but a clear in-plane heteroepitaxy was not observed, due to the limitation of a maximum annealing temperature of 700 °C. The estimated lattice parameters of the as grown $KFe_2As_2$ film are $a = 0.3924$ and $c = 1.374$ nm. This result indicates that the $c$-axis of the film slightly shrinks (–1 %) while its $a$-axis slightly expands (+2%), compared to those of the bulk from that of bulk ($c = 1.384$ nm, $a = 0.3841$ nm, Ref. 2), probably due to a relatively large in-plane lattice mismatch between $KFe_2As_2$ and MgO (+9%).

In conclusion, we developed a new solid-phase epitaxy technique to grow high-quality epitaxial films of compounds composed of alkali metals. It was demonstrated that for the $KFe_2As_2$ epitaxial films, the out-of-plane rocking curve width was decreased by an order of magnitude compared to previous results, and the in-plane XRD pattern completely overlapped with that of the MgO substrate. Since $KFe_2As_2$ is very air-sensitive and incompatible with conventional lithography techniques, we could not measure its SHC, which, however, can be achieved by developing a nanometer-size patterning technique for such highly air-sensitive materials. The solid-phase epitaxy technique employs a custom-designed alumina vessel and a sealing in Ar-filled stainless tube with a large amount of atmospheric powder (e.g., $KFe_2As_2$). This technique is a powerful tool for growing high-quality thin films composed of high vapor pressure elements.

Acknowledgments

This work was supported by the Ministry of Education, Culture, Sports, Science and Technology (MEXT) through Element Strategy Initiative to Form Core Research Center. H. Hi. was also supported by the Japan Society for the Promotion of Science (JSPS) Grant-in-Aid for Young Scientists (A) Grant Number 25709058, JSPS Grant-in-Aid for Scientific Research on Innovative Areas "Nano Informatics" (Grant Number 25106007), and Support for Tokyotech Advanced Research (STAR).

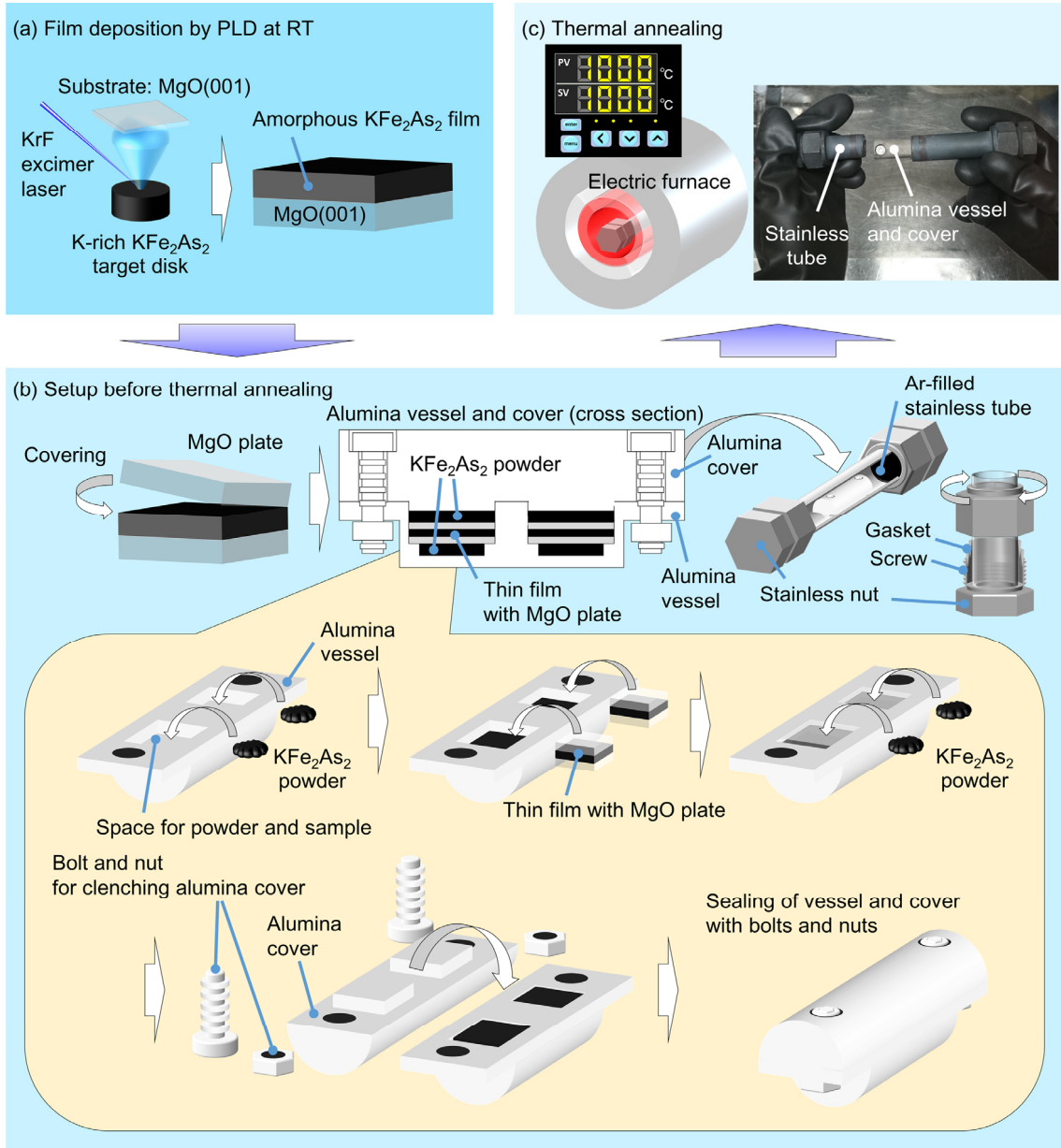

**Fig. 1.** Schematic of the experimental setup for a newly developed solid-phase epitaxy for growing KFe$_2$As$_2$ epitaxial films, via a post-deposition thermal annealing process. (a) Film-deposition process by PLD at room temperature (RT). (b) Setup before thermal annealing. The surface of the as-deposited amorphous film is covered with a fresh MgO

substrate. The film is then transferred to a custom-designed alumina vessel with small sample spaces. The films (maximum 2 samples) are covered with a large amount of stoichiometric $KFe_2As_2$ powder. Then the alumina vessel is tightly covered with a custom-designed alumina cover with commercially available bolts and nuts made of alumina to keep it gas-tight around the samples during high temperature thermal annealing. The tightly sealed alumina vessel is set in an Ar-filled stainless tube, both ends of which are sealed by stainless nuts and gaskets to keep it gas-tight. From (a) to (b), the sample is transferred directly from the preparation chamber of PLD (vacuum) to the glove box in pure Ar with a dew point of ca. $-100$ °C. The process in (b) is performed in the glove box. (c) Thermal annealing process in an electric furnace at 1000 °C (left). Then, the stainless tube is transferred to the glove box, and cut open by a tube cutter to pick out the sample (right photograph). The sample is not exposed to air throughout the whole process.

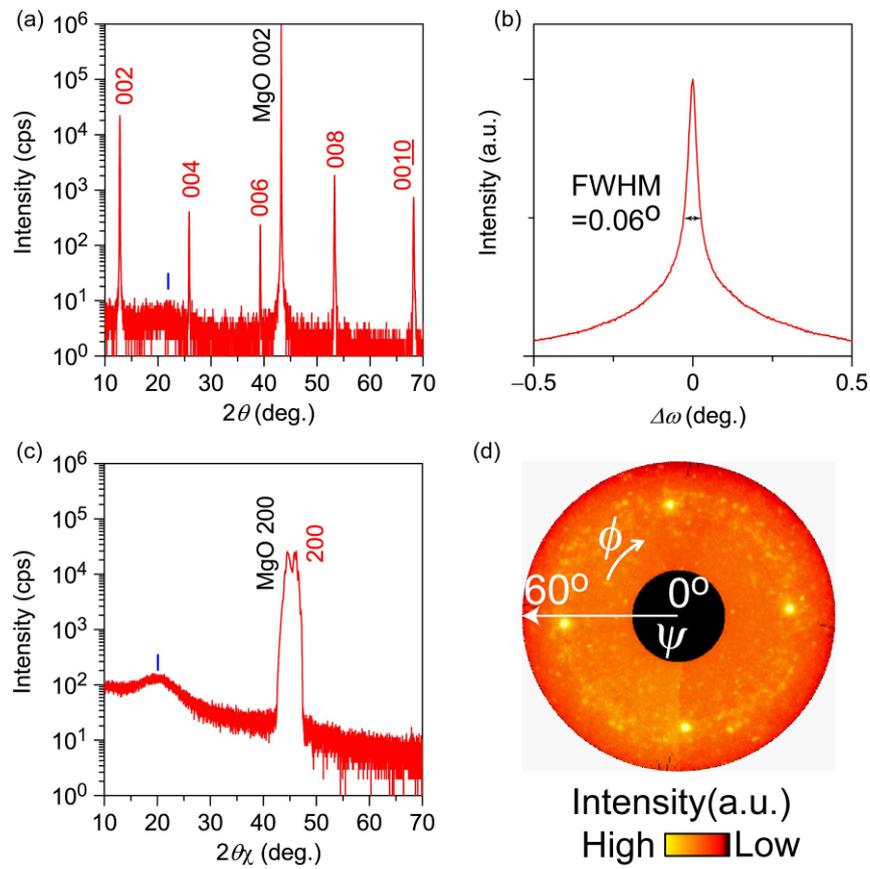

**Fig. 2.** XRD patterns of KFe$_2$As$_2$ epitaxial films grown at 1000 °C through a newly developed solid-phase epitaxy technique. (a) Out-of-plane $\omega$-coupled $2\theta$ synchronous scan pattern. (b) Out-of-plane rocking curve of the 002 diffraction. (c) In-plane $\phi$-coupled $2\theta\chi$ synchronous scan pattern. (d) Pole figure of the asymmetric 103 diffraction. The vertical bars in (a) and (c) indicate the halo patterns from the O-ring sealed sample carrier filled with dry Ar gas.

Supplementary data for "Novel solid-phase epitaxy for multi-component materials with extremely high vapor pressure elements: An application to KFe$_2$As$_2$"

Taisuke Hatakeyama,[1] Hikaru Sato,[1] Hidenori Hiramatsu,[1,2] Toshio Kamiya,[1,2] and Hideo Hosono[1,2]

[1] Laboratory for Materials and Structures, Institute of Innovative Research, Tokyo Institute of Technology, Japan

[2] Materials Research Center for Element Strategy, Tokyo Institute of Technology, Japan

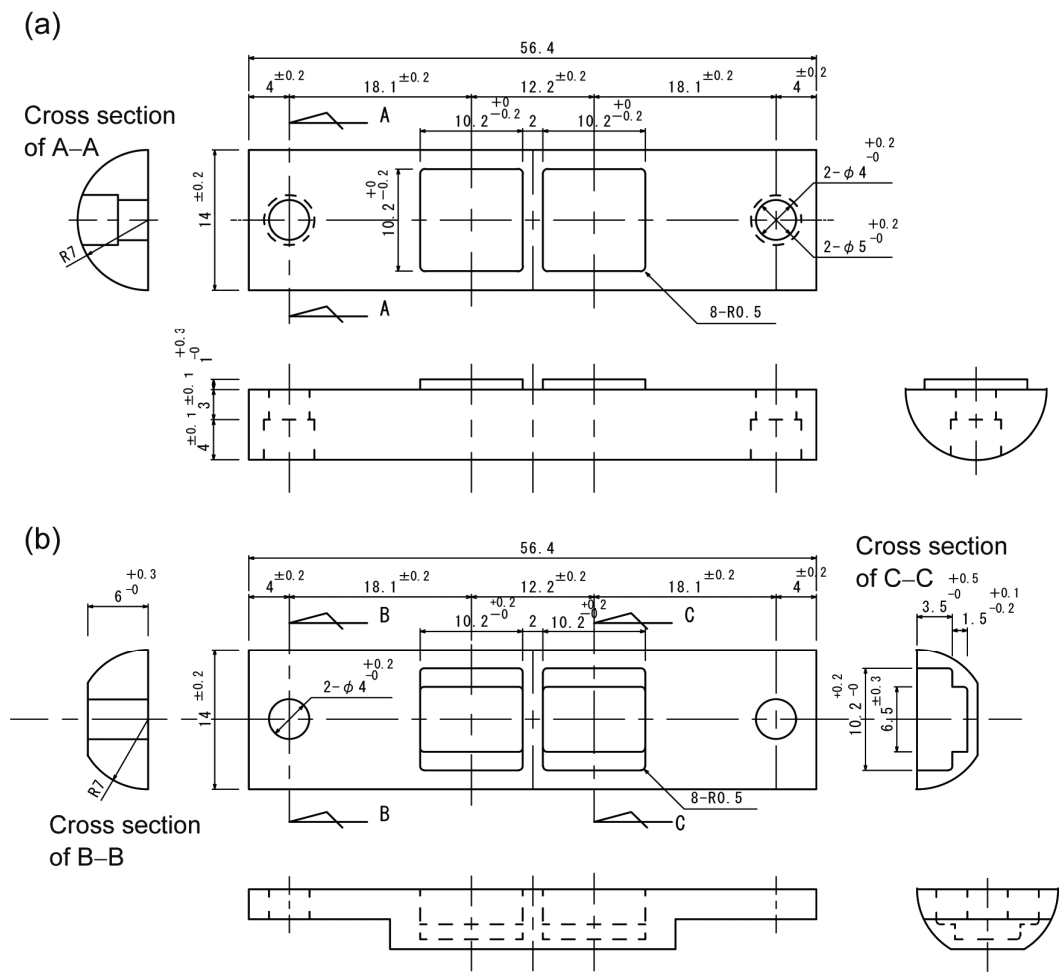

**Figure S1.** Drawings of the custom-designed cover (a) and vessel (b) made by alumina. The unit is in mm. Small characters with signs indicate process tolerances.